\documentclass[a4paper,fleqn]{cas-sc}

\usepackage[authoryear,longnamesfirst]{natbib}

\def\tsc#1{\csdef{#1}{\textsc{\lowercase{#1}}\xspace}}
\tsc{ICDL}
\tsc{ABS}
\tsc{TAN}
\tsc{HR}

\begin{document}
\let\WriteBookmarks\relax
\def\floatpagepagefraction{1}
\def\textpagefraction{.001}
\shorttitle{Orientation of Swimming Cells with Annular Beams}
\shortauthors{ICD Lenton et~al.}

\newcommand{\ecoli}{\emph{E. coli}}

\title [mode = title]{Orientation
    of Swimming Cells with Annular Beam Optical Tweezers}                      
\author{Isaac C. D. Lenton}[orcid=0000-0002-5010-6984]
\cormark[1]
\ead{uqilento@uq.edu.au}

\credit{Software, Investigation, Writing - Original Draft}

\author{Declan J. Armstrong}[orcid=0000-0001-6602-9832]
\ead{declan.armstrong@uq.net.au}

\credit{Investigation}

\author{Alexander B. Stilgoe}[orcid=0000-0002-9299-5695]
\ead{stilgoe@physics.uq.edu.au}

\credit{Writing - Review \& Editing, Supervision}

\author{Timo A. Nieminen}[orcid=0000-0003-3055-7275]
\ead{timo@physics.uq.edu.au}

\credit{Funding Acquisition, Writing - Review \& Editing, Supervision}

\author{Halina Rubinsztein-Dunlop}[orcid=0000-0002-8332-2309]
\ead{halina@physics.uq.edu.au}

\credit{Funding Acquisition, Writing - Review \& Editing, Supervision}

\address{School of Mathematics and Physics, The University of Queensland, Brisbane, Queensland 4072, Australia}

\cortext[cor1]{Corresponding author}

\begin{abstract}
Optical tweezers are a versatile tool that can be used to
manipulate small particles including both motile and non-motile
bacteria and cells.
The orientation of a non-spherical particle within a beam depends
on the shape of the particle and the shape of the light field.
By using multiple beams, sculpted light fields or dynamically
changing beams, it is possible to control the orientation of certain
particles.
In this paper we discuss the orientation of the rod-shaped
bacteria \emph{Escherichia coli} (\ecoli{}) using dynamically
shifting annular beam optical tweezers.
We begin with examples of different
beams used for the orientation of rod-shaped particles.
We discuss the differences between orientation of
motile and non-motile particles, and explore
annular beams and the circumstances when they may be beneficial
for manipulation of non-spherical particles or cells.
Using simulations we map out the trajectory the \ecoli{} takes.
Estimating the trap stiffness along the trajectory gives us an insight
into how stable an intermediate rotation is with
respect to the desired orientation.
Using this method, we predict and experimentally verify
the change in the orientation of motile \ecoli{} from
vertical to near-horizontal with only one intermediate step.
The method is not specific to exploring the orientation of
particles and could be easily extended to quantify
the stability of an arbitrary particle trajectory.
\end{abstract}




\textbf{Pre-print of:}
\begin{quote}
    I. Lenton, et al., 
    Orientation of swimming cells with annular beam optical tweezers,
    Optics Communications,
    2019, 124864, ISSN 0030-4018,
    \url{https://doi.org/10.1016/j.optcom.2019.124864}.
    
\copyright~2019. This manuscript version is made available under the CC-BY-NC-ND 4.0\\
license \url{http://creativecommons.org/licenses/by-nc-nd/4.0/}.
\end{quote}

\begingroup
\let\pagebreak\relax
\maketitle
\endgroup



\section{Introduction}

In 2018 half of the Nobel prize in physics was awarded to
Arthur Ashkin for the invention of optical tweezers 
\citep{Ashkin1986May} and their
application to the study of biological systems.
Optical tweezers consist of one or more laser beams which can be
used to apply pico-newton scale forces to small particles in order
to trap them in three dimensions.
Since the demonstration of three
dimensional trapping by Ashkin in 1986 and a first demonstration of
using it in biological systems in 1987 \citep{Ashkin1987Mar},
optical tweezers have been studied
in numerous research labs around the world and used for broad studies of
biological systems, reaching on one hand single molecule
detection \citep{Lang2004Oct}
and on the other hand trapping of very large objects deep in
living tissue \citep{Favre-Bulle2017Sep}.
Optical tweezers can be used to trap spherical particles, as well
as a range of non-spherical particles, either in a single beam or
with multiple beams to orientate the particle in a desired direction.

One example of a non-spherical particle that we are
interested in
is the rod-shaped bacteria
\emph{Escherichia coli} (\ecoli{}) \citep{Berg2004}.
This bacteria is of interest as a model organism which can be
used to study, amongst other things,
micro-scale fluid flow and swimming near surfaces.
Using optical tweezers it is possible to hold motile
\ecoli{} by balancing the motility force with an
equal but opposing force from the optical tweezers.
By measuring the scattered optical tweezers beam,
it is possible to get an accurate measurement for the
swimming force \citep{Bui2018Jul}.
Further, by tracking
the particle's position,
it is possible to simultaneously measure the
velocity enabling studies of the
relationship between swimming velocity and
swimming force.
In order to facilitate these studies, we need to be
able to manipulate \ecoli{} in order to orientate
the particle in a desired direction.

Some cells can be directly manipulated using holographic
optical tweezers (HOT).
For example, \citet{Horner2010Jul} trap and orientate
multiple \emph{Bacillus subtilis} (a $3\mu$m bacterium similar to \ecoli{})
simultaneously using HOT with two traps, one at each end of
the bacterium.
The traps can be gradually moved in order to rotate the
particle or change the particle's position.
In another experiment, \citet{Carmon2011Jan} rapidly scan a single beam
between the two end points of the bacterium.
The advantage of scanning the beam in this way avoids interference between
the optical traps, allowing for tighter confinement
in the axial direction.
Particles can also be aligned using structured light fields
or other more specialised approaches.
For example, optical tweezers formed at the tip
of an optical fibre have been used to create structured light fields
for orientating \ecoli{} \citep{Huang2015Dec}.
HOT can be used to manipulate several particles at once or
rotate larger structures such as micro-rotors, which in turn
can be used to generate fluid flows for
indirectly manipulating particles \citep{Butaite2019Mar}.

Our present goal is to align motile \ecoli{} perpendicular to the
beam axis to enable us to study how these cells behave in
certain environments.
Although \ecoli{} and similar rod-shaped bacteria have been previously
orientated using  scanned beams
and HOT
\citep{Horner2010Jul, Carmon2011Jan}, we found
it can be difficult to reproduce these results using
our existing experimental setup.
Factors such as the numerical aperture, aberration or greater
motility of the bacteria compared to these previous studies make the
experiment difficult to perform.
We previously had success orientating several-days old \ecoli{}
(unhealthy, less motile) using HOT and line shaped traps.
We were
unable to reproduce the results with healthier and more motile \ecoli{}.
We also explored using other shaped beams including tug-of-war tweezers,
which have been demonstrated for stretching elongated
cells \citep{Lamstein2017Mar}.
While some of these approaches worked in simulations, it can
be difficult to realise these traps in a lab experiment.
Aberrations and an insufficient numerical aperture can reduce
the trapping effectiveness of intricate structured light fields.

In this work, we describe the use of annular beams to hold \ecoli{}
and dynamically change the potential in order to align the
particle to a desired orientation.
Annular beams have a simple structure which has been previously
observed to reduce back reflection
and thus can improve axial optical trap depth---one of the
problems in our earlier experiments.
We use a spatial light modulator to generate and dynamically
change the position of the annular beams.
Using simulations, we are able explore the angular trap stiffness
along the path the \ecoli{} takes when we shift the beams.
We investigate the number of intermediate patterns required
to change the \ecoli{} orientation from vertical to horizontal
in order to increase the transition speed on devices with a
finite frame rate.
Our method is simple and robust since it only requires a device
capable of displaying a few discrete patterns and doesn't depend
on specialised high speed devices.
The paper is split into two main parts: a background section
which provides an overview of optical trapping and the generation
of annular beams, and a results section which describes our
investigations targeted at orientation of motile \ecoli{}.

\section{Background}

The principal behind optical tweezers is the transfer
of momentum from one or
more laser beams to a particle.
Scattering and absorption by the particle leads to a change in the
laser beams momentum, resulting in a corresponding force on the particle.
The force transferred to the particle depends on the beam power,
the amount of scattering and the speed of light in the medium.
The amount of scattering in-turn depends on the particle size and shape,
the relative refractive index of the particle in the surrounding medium,
as well as the overlap between the beam (shape) and the particle (shape).
The optical force is often separated into two components: the
gradient force and the scattering force.
The gradient force is proportional to the gradient of the
electric field and is either attractive or repulsive depending
on the refractive index of the particle relative to the surrounding
medium.
Particles with a refractive index greater than the surrounding
medium will be attracted by the gradient force towards more
intense regions of the beam, while
particles with a lower refractive index will be repulled.
The scattering force arises from light reflected or absorbed
by the particle, and often acts to push the particle along the
beam axis.
For particles with a higher refractive index than the surrounding
medium, stable trapping in three dimensions requires overcoming
the scattering force.
This can be achieved, for instance, by using
counter-propagating beams \citep{Ashkin1970Jan},
balancing the scattering force with another force such as
gravity \citep{Ashkin1971Oct},
or by using a tightly focused beam such that the gradient force
overcomes the scattering force \citep{Ashkin1986May}.

\begin{figure}
    \centering
    \includegraphics{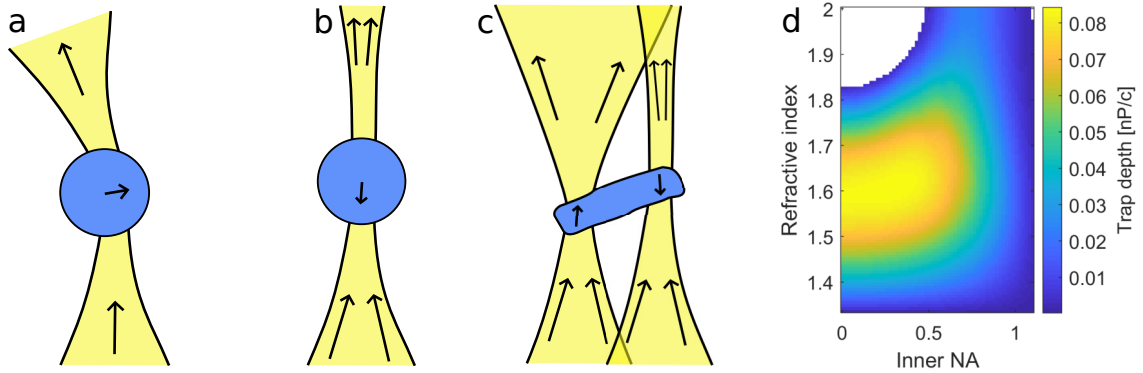}
    \caption{(a--c) Illustrations showing the optical force
    on dielectric particles due to the transfer of momentum
    from the light beam to the particle.
    (a) and (b) show optical forces on a spherical
    particle trapped in a Gaussian-like beam.
    (c) shows a elongated particle in a dual beam optical trap.
    (d) demonstration of how annular beams can be used to
    improve trap depth for trapping of a spherical particle
    (radius $R \approx 0.46\lambda_0$)
    in water (refractive index $n = 1.33$).
    As light is removed from the centre of the beam
    (Inner NA $\rightarrow 1$),
    the range of particles which can be trapped increases.}
    \label{fig:trapping-overview}
\end{figure}

The simplest optical traps typically involve using a tightly
focused laser beam, often a beam with a Gaussian profile,
to trap and manipulate particles.
For spherical particles trapped in a Gaussian beam,
the gradient force typically results in the particle being
trapped at the beam focus or, if the particle is strongly scattering, 
slightly downstream of the focus.
A small dielectric particle can be thought of as a small lens.
When the particle moves through the beam it will change the direction
and collimation of the beam.
The particle will experience a corresponding force, opposing the
change in momentum of the beam, as illustrated in
figure~\ref{fig:trapping-overview}~(a--b)
for a particle with a refractive
index higher than the surrounding medium.
The optical force can be increased by changing the
refractive index contrast between the particle and medium,
increasing the power, or by changing the beam phase/amplitude
distribution.
In some circumstances, increasing the refractive index
contrast can improve trapping; however, the corresponding
increase to the scattering force often leads to the particle
no longer being stably trapped in three dimensions.
By using structured light beams, such as annular beams, it is
possible to reduce the scattering force and improve the
trap quality.

Trap quality can be measured in a number of ways
including trap stiffness, which is a measure of how
steep the gradient is around the trap centre; and
trap depth, a measure of how much force can be applied
before the particle escapes.
In this paper, we define the trap depth as the minimum
of the peak restoring force/torque in a particular
direction, i.e., the maximum force or torque that can
be applied to the particle before it escapes the trap.
The trap depth has the units of force.
We use the dimensionless quantities $nP/c$
and $P/\omega$ for optical force and torque, where $n$
is the refractive index of the medium, $P$ is the beam power
and $c$ is the speed of light in vacuum.
These quantities give the force in units of $n\hbar k$
per photon and torque in units of $\hbar$ per photon
\citep{Nieminen2014Oct}.
Similar quantities can be defined for 
non-spherical particles
in terms of the torque aligning the particle to a particular
direction.

Figure~\ref{fig:trapping-overview}~(d)
shows how trap depth for a spherical particle in an annular beam
changes as a function particle refractive index and
beam shape (in this case
the inner numerical aperture (NA) of the annular).
Annular beams may refer to beams where the centre has been
removed in the far-field, or the term may refer to a beam with
a phase discontinuity at the centre, creating a doughnut shape
in the near-field.
In this paper we use the term annular beam to refer to
beams where the central portion has been removed
in the far-field.
This type of annular beam can be described by two angles
(numerical apertures)
for the interior and exterior ring radii.
For reflective or absorbing particles, annular beams
can improve trapping by reducing the amount of
light in the centre of the beam contributing to the scattering
or absorption force \citep{Ashkin1992Feb, Padgett2011May}.
The top left corner of figure~\ref{fig:trapping-overview}~(d)
shows a region where
a high contrast particle cannot be trapped.
By using an annular beam, a range of particles with a much
larger variety of refractive indices can be trapped
and the trap depth
is improved for certain refractive-index/beam combinations.
The motivation for this current study was to understand if
a similar improvements could be seen for the orientation of
rod-shaped particles perpendicular to the beam axis.

For non-spherical particles, it is often desirable to be able to
control not just the position but also the orientation.
Small elongated particles, such as \ecoli{},
held in a Gaussian beam, tend to align either along the beam axis
or along the polarisation axis, depending on the size
and aspect ratio \citep{Simpson2011Nov, Cao2012Jun}.
If the particle is sufficiently large,
it may be possible to grab different parts of the
particle with multiple traps, as illustrated in 
figure~\ref{fig:trapping-overview}~(c).
Or, it may be possible to use beam shaping to generate
a light field which orientates the particle in the
desired orientation.
In any case, there are frequently multiple stable
equilibria which the particle may become aligned to.
For example, in the twin beam case shown in
figure~\ref{fig:trapping-overview}~(c), the particle
could become aligned in either of the beams individually.

\begin{figure}
    \centering
    \includegraphics[width=\textwidth]{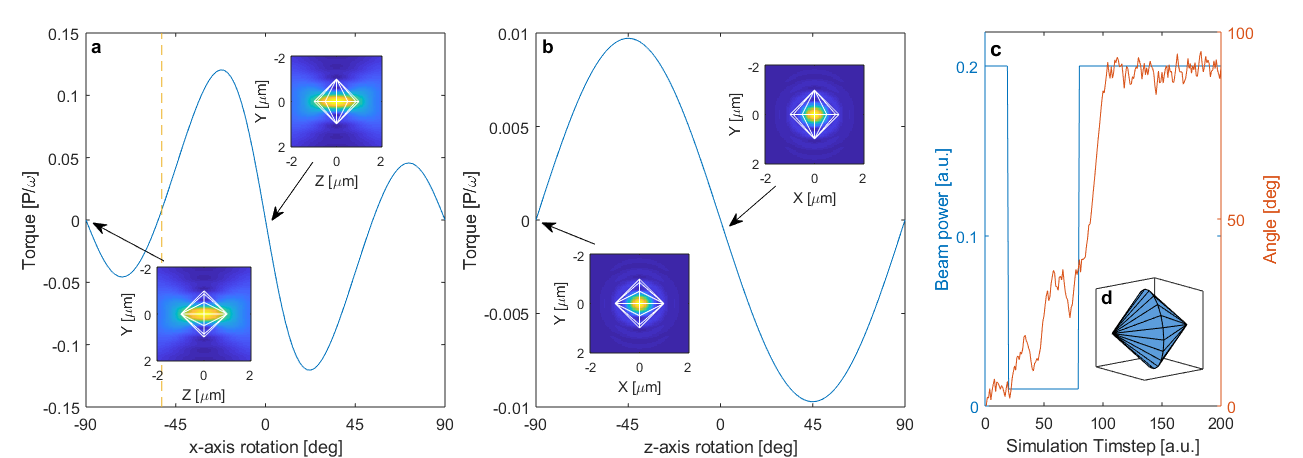}
    \caption{(a-b) torque-rotation plots for a particle (d) with
    multiple stable angular equilibria.
    The dashed line in (a) marks the unstable equilibrium
    between the two stable traps.
    Insets show particle orientation in the beam.
    (c) shows the beam power and the angle between the
    particle axis and beam axis from Movie~1.
    (d) shows the particle used in these simulations, a
    reminiscent of two cones stacked on top of each other.
    }
    \label{fig:nonmotile}
\end{figure}

When the particle size approximately matches the size of the beam
or features of the shaped beam, there are often multiple stable
equilibria.
Alignment of these particles can be achieved by switching
between the different equilibria.
For certain non-motile particles, such as certain crystals or
bacteria spores, particles will naturally drift between different
equilibria if the temperature is high enough or the trap power
is low enough.
Thus, is is possible to switch between
equilibria by simply lowering the power of the laser,
letting the particle diffuse through Brownian motion, and
increasing the power once the desired orientation has been
achieved; as shown for a crystal-like dielectric particle
in  Figure~\ref{fig:nonmotile} and Movie~1.
Movie~1 demonstrates this method for changing
the orientation between two stable equilibria, separated
by an unstable equilibrium at $\sim{}\!38^\circ$.
The beam power is lowered to allow the
particle to rotate through Brownian motion.
Once the particle has passed $\sim{}\!38^\circ$,
corresponding to the unstable equilibrium
marked in Figure~\ref{fig:nonmotile}~(a),
the trap power is raised and the particle falls into the
new equilibrium.
The positions of the stable and unstable equilibria, and the
corresponding times when beam power must be raised/lowered,
depend strongly on the particle size/shape.

Trapping of live cells including bacteria is more difficult.
If the power is too high, cells can be damaged by absorption
and subsequent heating \citep{Zhang2008Apr, Neuman2004Sep}.
In addition, many living cells are also motile, some moving at
speeds of microns per second.
If the power is too low, these particles tend
to escape from the trap.
This makes it difficult to move the particle between different
equilibria using the previously described method, since lowering
the power can lead to the particle
rapidly swimming out and away from the trap.
Furthermore, when trapped, motile particles are able to explore
a greater range of the optical potential, often escaping traps
that would hold similarly shaped non-motile particles.
By using sculpted light beams, it is possible to create traps
which strongly confine a particular shaped particle to
a specific position and orientation.
The orientation and position can be changed by dynamically
changing between different shaped beams.

\begin{figure}
    \centering
    \includegraphics[width=0.7\textwidth]{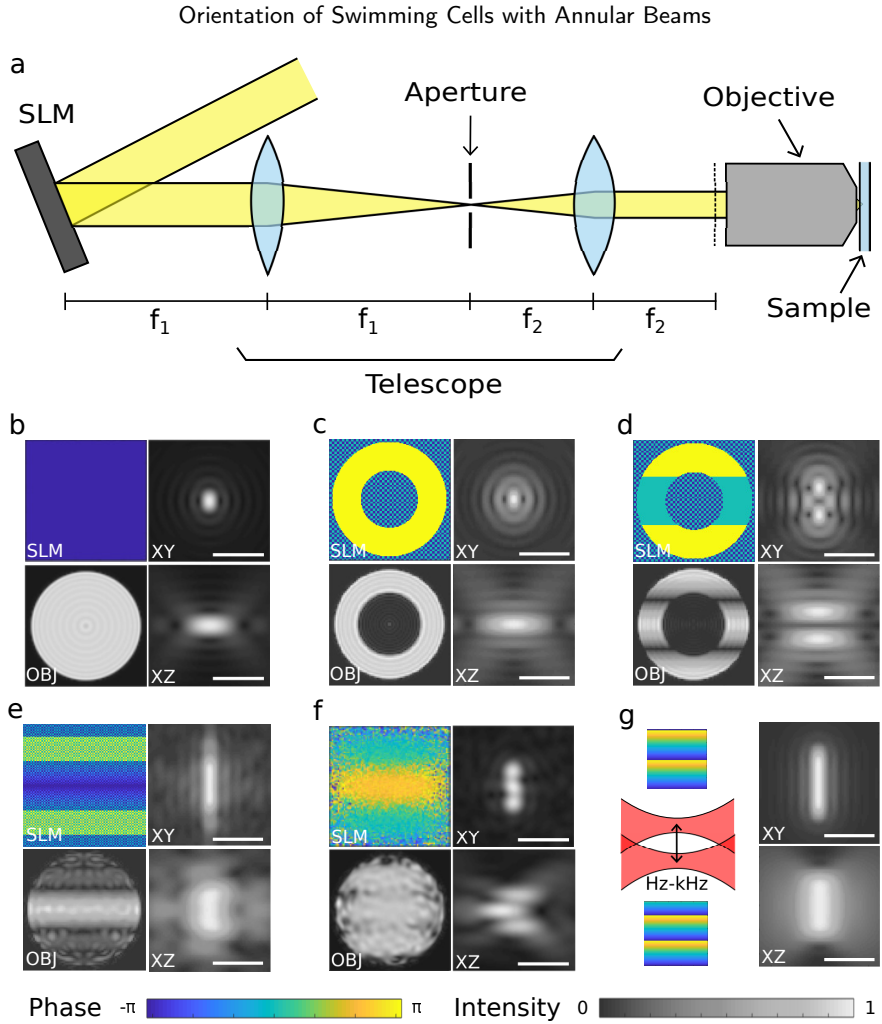}
    \caption{(a) overview of optical tweezers system including
    a spatial light modulator (SLM) for phase modulation.
    The aperture is used to remove light scattered to
    high angles.
    $f_1$ and $f_2$ are the focal lengths of the lenses
    forming the telescope.
    (b--g) simulations of different kinds of beams
    and the corresponding phase at the SLM plane:
    (b) uniform phase and illumination,
    (c) annular beam, (d) twin annular beam,
    (e) line trap generated using sinc pattern,
    (f) line trap generated using Gerchberg-Saxton
    algorithm, (g) line trap generated by scanning
    the beam.
    OBJ shows the light intensity at the objective
    back-focal plane, marked by the dotted line in (a).
    XZ and XY show the light near the focal plane,
    the scale bar shows $\sim 2\mu$m.
    All beams have linear polarisation.
    The aliasing effect in (c--e) SLM is an artefact of rendering
    the 512x512 pixel checkerboard pattern at a lower resolution
    in the figure.}
    \label{fig:overview}
\end{figure}

There are a number of methods for generating different shaped
light fields including using multiple beams, modulating the
phase or amplitude of a single beam, or rapidly
scanning a single beam between multiple positions.
Multiple traps can be generated using different lasers,
beams with different polarisation, or with diffractive
optical elements to split the light between multiple traps.
Devices such as the digital micro-mirror
device or liquid crystal spatial light modulator (SLM)
can be used to rapidly change the intensity or phase
of the beam.
These devices can be fast, with some devices operating at
kHz speeds \citep{Gauthier2016Oct, Stuart2018Feb};
they can be used to create multiple holographic optical
tweezers (HOT) for manipulating multiple particles
simultaneously \citep{Dholakia2011May, Padgett2011May};
and they can also be used to change the shape of the
beam in order to better match the shape of the
particle and improve trap stiffness/depth
\citep{Woerdemann2013Nov, Roichman2006Jun}.
In this paper we use a phase-only SLM to
modulate the incident beam.
The SLM is imaged onto the back focal plane of the microscope
objective, as shown in figure~\ref{fig:overview}~(a).
A telescope can be used to reduce or enlarge the beam
in order to fill (or over/under-fill) the microscope
back aperture.
An aperture between the telescope lenses allows additional spatial
filtering of the beam, this can be useful for achieving
amplitude control with a phase-only device.


Figure~\ref{fig:overview}~(b--g) shows different kinds of
beams simulated using the optical tweezers toolbox (OTT)
\citep{Nieminen2007Jul, Lenton2019Jul}.
Optical tweezers systems typically use high NA
objectives and the resulting focused fields are non-paraxial.
In order to simulate the focused fields and calculate the
optical forces/torques we used the point matching method
to match the paraxial far-field to the vector spherical
wave-function expansion in the far-field
\citep{Nieminen2003Jun}.
We simulated a water immersion objective (NA=1.2) and
assumed a $\sin\theta$ mapping between SLM pixel coordinates
and focusing angle.
Figure~\ref{fig:overview}~(b) shows a beam with uniform phase
and uniform amplitude at the back aperture of the objective.
This beam can be converted into an annular beam by simply
removing the central portion of the beam, 
as shown in figure~\ref{fig:overview}~(c).
An elongated particle will align in these beams either along
the beam axis or along the polarisation axis, depending on
size/shape.

Our present interest is in orientating \ecoli{}.
In order to do this, we have investigated a number of
different kinds of beams.
Figure~\ref{fig:overview}~(d--g) show different kinds of
beams intended to align rod-shaped particles perpendicular
to the beam axis.
The simplest configuration is dual beam optical tweezers,
which have previously been demonstrated for trapping
\ecoli{} \citep{Horner2010Jul}.
These can be generated by simply superimposing the diffraction
pattern for each beam (a linear phase grating controlling the
position of each beam).
Figure~\ref{fig:overview}~(d) shows dual annular beam optical
tweezers, the phase pattern is simply the dual beam optical
tweezers phase pattern with the centre removed.
We also explored the stripe (or line) beam \citep{Roichman2006Jun},
shown in figure~\ref{fig:overview}~(e).
Non-motile particles would align to the beam but motile
particles would swim out the ends of the beam.  We assumed
this was because of the weaker trap strength along the line.
Figure~\ref{fig:overview}~(f) shows a beam generated using
the Gerchberg-Saxton algorithm \citep{Gerchberg1971Nov}.
The beam has sharp features in the XY plane, however the
axial intensity shows large lobes.
In simulations, we found motile \ecoli{} tend to align
vertically in these kinds of lobes and don't trap in the
desired orientation.
Figure~\ref{fig:overview}~(g) shows a scanned beam.
Theses beams have been previously shown to be able to hold
\ecoli{} \citep{Carmon2011Jan} however our current experimental
system doesn't support creating beams with this method.

Simulations suggested that the dual (annular) beams should
hold \ecoli{} horizontally, however our initial efforts at
orientating \ecoli{} with these beams were unsuccessful.
When we simulated particles swimming into the trap, we found
that they would tend to align in a single beam and not reach
the centre of the trap.
In the following section we discuss the use of annular
beams and dynamically changing the SLM pattern for the
orientation of \ecoli{}.

\section{Orientation of motile \ecoli{}}

This section is split into three parts: generation of dual annular
beam optical tweezers using an SLM, orientation of motile
\ecoli{} using dynamic potentials and an overview of a method
for characterising the \ecoli{} trajectory.

\subsection{Generation and Control of Annular Beams}

Annular beams are characterised by a distinctive darkened patch in their
centre far from a focus and can be created in a number of ways including
using a SLM (see figure~\ref{fig:overview}~(c--d)), with a fixed
mask \citep{Oliveira2018Dec, Dear2012Oct}, or with a pair of
axicons \citep{Lei2013Mar}.
In this section we investigate the use of annular beams on
rod-shaped particles to determine if there is a similar improvement
for angular trap stiffness.

We use a computer controlled SLM to create the annular beams used in the experiments.
We control the annular beam position by applying a linear phase function for
transverse displacement and a parabolic phase function for axial displacement,
\begin{eqnarray}
    \phi_\mathrm{lens}(x, y) &=& \frac{x^2 + y^2}{2R} \\
    \phi_\mathrm{linear}(x, y) &=& Dx
\end{eqnarray}
where $D$ and $R$ control the magnitude of displacement for the beam.
To rotate the \ecoli{}, we combine the phase patterns for each
beam using
\begin{equation}
    \phi_\mathrm{twin} = \arg\left\{
        e^{\mathrm{i}\phi_\mathrm{linear}\sin\theta + \mathrm{i}\phi_\mathrm{lens}\cos\theta} + 
        e^{-\mathrm{i}\phi_\mathrm{linear}\sin\theta - \mathrm{i}\phi_\mathrm{lens}\cos\theta}\right\}
\end{equation}
this produces a double spot pattern with elliptical (circular for
calibrated $D$ and $R$) trajectory as a function of angle $\theta$.
The choice of $D$ and $R$ affects the trap stiffness along the particle axis;
these parameters can be adjusted to give good trap stiffness in the vertical and
horizontal particle orientations.

An annular beam is generated by removing the inner central portion of a Gaussian or flat-top beam.
With a amplitude-type SLM, such as a digital micro-mirror
device, the required mask can be generated by simply
setting the output in the central region to zero.
We use a phase-only SLM, which can not directly control
the beam amplitude.
Instead, we introduce a second pattern to scatter light
to other locations in the far-field.
We use a checkerboard pattern to scatter light to large angles
outside our optical path \citep{Wong2008Feb, Stilgoe2016May}.
A iris or aperture can be used to remove this light, as
shown in figure~\ref{fig:overview}~(a).
The resulting pattern is
\begin{equation}
    \phi_\mathrm{annular}(x, y) = \left\{ \begin{array}{cc}
        \phi_\mathrm{twin} & r_1 < \sqrt{x^2 + y^2} < r_2  \\
        \phi_\mathrm{checker} & \text{otherwise}
    \end{array} \right.
\end{equation}
where $r_1$ and $r_2$ are the inner and outer radius of the
annular beam.
In our experiments we set $r_2$ to approximately the same size
as the radius of the objective back aperture.
To implement these functions and control the SLM we use our
soon-to-be released beam shaping toolbox, OTSLM.
An example phase pattern and simulated far-field for one of these
double-spot annular beams is shown in 
figure~\ref{fig:overview}~(d). 

\begin{figure}
    \centering
    \includegraphics[width=0.6\textwidth]{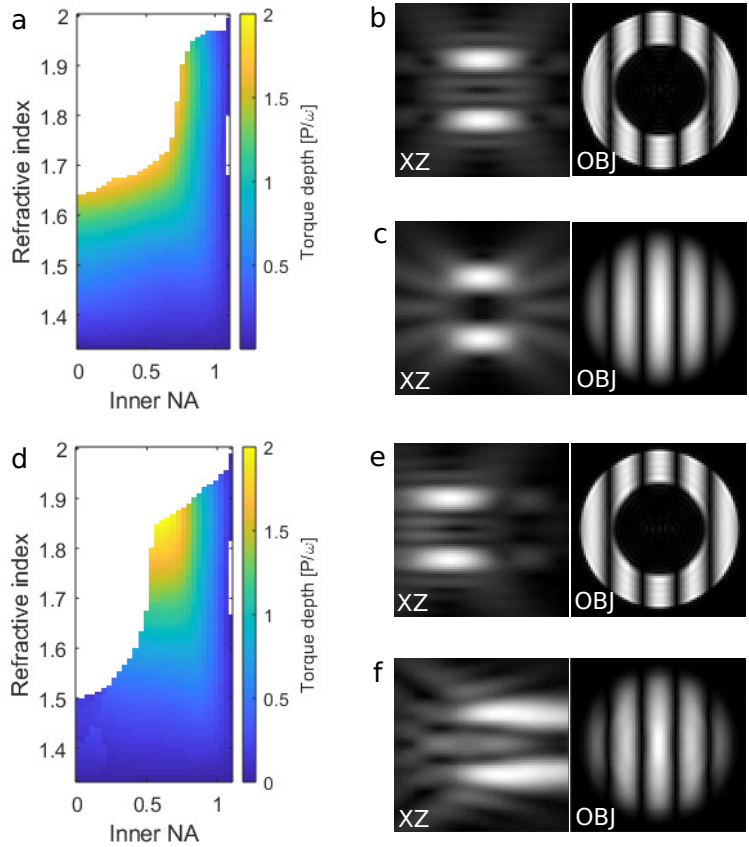}
    \caption{Exploration using dual annular beam optical tweezers
    (separation $1.8\mu$m)
    to trap a rod-shaped particle (length $3\mu$m, radius $0.5\mu$m)
    horizontally.
    (a) angular trap depth for rod-shaped particle,
    (b--c) annular and Gaussian beam near-field (XZ)
    and far-field (OBJ) intensity.
    (d--f) show how the previous results change
    with the introduction of a spherical aberration.}
    \label{fig:annularbeams}
\end{figure}

To understand effect of using these beams on the
angular trap depth, we used the optical tweezers
toolbox to simulate the beams \citep{Nieminen2007Jul, Lenton2019Jul}.
We choose to explore how changing the inner radius of the
annular beam while keeping the beam power fixed affects the
trap depth.
Figure~\ref{fig:annularbeams}
shows the angular trap depth for rod-shaped
particles in twin annular beams with and without
spherical aberration.
We first determined if the particles could be trapped and if the
particle could be orientated horizontally between the two annular beams.
For rod-shaped particles in the twin annular beams, we observed
very little change for low refractive indexes like \ecoli{}
in water; in some cases the trap depth was reduced
by using annular beams.

By using annular beams, there is an improvement to the range of high
index particles which can be trapped, this would be useful, for instance,
trapping rod-shaped micro-organisms or spores in air or vacuum.
When certain spherical aberrations are added to the beams, both the range of
particles that can be trapped and the trap depth are significantly reduced
for non-annular beams but appear to improve for annular beams.
Investigating how these beams look with spherical aberration, we see
how the annular beams maintain their general shape for small
deviations, while Gaussian-like beams become more elongated
in the axial direction (figures~\ref{fig:annularbeams}~(b--c, e--f)).
In some cases, these spherical aberrations may enhance trap stiffness
or create additional trapping equilibria \citep{Stilgoe2011Dec}.
While annular beams may not provide an improvement in general, it is important
to consider the size of the annular beam as well as the direction of the
spherical aberration.

These results, although interesting, are not particular useful for
trapping of \ecoli{} in water.
In systems with strong spherical aberrations, it is often possible
to correct for the aberration using
an SLM or phase mask\citep{Itoh2009Aug, Wulff2006May}.
However, using an annular beam can be simpler than implementing
one of these methods.
Annular beams would be more suitable for trapping of high contrast
particles, such as trapping cells in air or
vacuum \citep{Pan2011Apr}.

\subsection{Orientating Motile \ecoli{} with Dynamic Potentials}

Our simulations showed that one of the difficulties  with
orientating motile \ecoli{} was with loading them into the dual
beam trap.
The \ecoli{} would often align with one of the annular beams
rather than aligning between the two.
This difficultly was encountered for both annular and Gaussian
dual beam tweezers.
Although rod-shaped bacteria have been previously orientated using this
method \citep{Horner2010Jul}, we found it difficult to reproduce
these results in our experiment using \ecoli{},
perhaps due to increased motility or the shorter length of the bacteria.
To orientate the \ecoli{} more reliably, we found it better
to start with the \ecoli{} aligned to the beam axis
before attempting to change the orientation.
By gradually changing the SLM pattern, we were able to orientate
the particles in the desired direction.

In our experiment, we trap motile \ecoli{} using HOT; the
setup has been previously described in \citep{Kashchuk2019Apr},
and here we only give a brief summary.
We use a 1064 nm fiber laser (YLR-10-1064-LP, 10W, IPG Photonics) focused
by a high numerical aperture
objective (Olympus UPlanSApo $60\times$, water immersion, 1.2 NA).
For beam shaping, we use an SLM
(Meadowlark Optics, 512x512 HSP512L, high-speed SLM)
imaged onto the back focal plane of the objective (as shown in
figure~\ref{fig:overview}).
The \ecoli{} is trapped in a buffer solution
(consisting of potassium phosphate, EDTA and KCI), with a refractive
index similar to water, between two microscope coverslips spaced
with double-sided tape ($\sim{}$0.1--0.2mm thick).

\begin{figure}
    \centering
    \includegraphics[width=\textwidth]{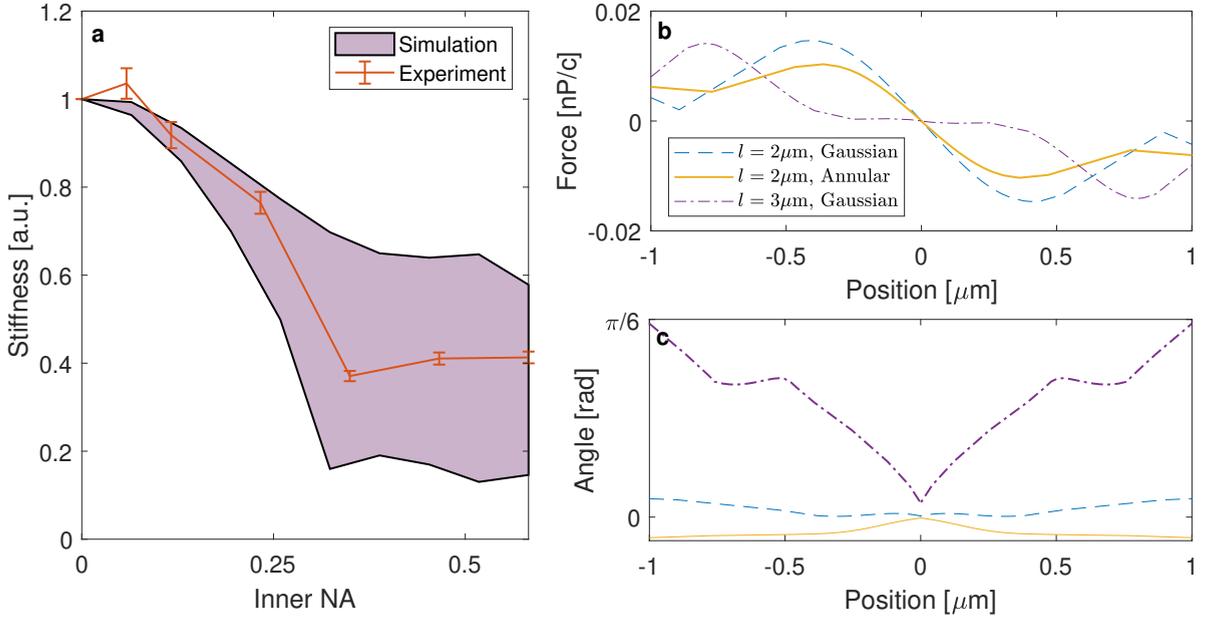}
    \caption{Exploration of trap properties for \ecoli{} in
    twin annular beams.  (a) shows
    numerical modelling and experimental measurements of the trap
    stiffness in the direction of trap separation for motile \ecoli{}
    as a function of inner beam radius for fixed SLM illumination.
    The shaded region shows the range of predicted trap stiffness
    for different ratios of the trap power and
    \ecoli{} motility force.  Approximate values for the
    \ecoli{} shape (length $l = 2\mu$m and radius $r = 0.5\mu$m)
    and refractive index ($n = 1.36$) were used for
    numerical simulations.
    (b) Simulation of the measured
    force for Gaussian-like (Inner NA $= 0$) and
    Annular (Inner NA $= 0.58$) beams with different
    \ecoli{} lengths ($l$).
    (c) angle of \ecoli{}
    relative to the plane transverse to the beam axis for the
    beams in (b).}
    \label{fig:stiffness}
\end{figure}

We started with the \ecoli{} trapped in a single beam and then
displayed a sequence of patterns to gradually move the traps from
both being aligned with the beam axis to both being separated
transverse to the beam axis.
To trap the \ecoli{}, we tried different parameters controlling
the inner annular radius, linear grating periodicity and lens grating spacing.
The separation of the two traps (i.e., the linear and spherical grating
parameters) were roughly determined by the length of the particle,
as previously noted by \citep{Catala2017Feb}.
To measure the quality of the trap, we used measurements of the
force in the direction of trap separation to determine trap
stiffness.
Figure~\ref{fig:stiffness}~(a) shows results for different inner
radius of the annular beam; the incident power on the SLM was
held fixed.
Trap stiffness was determined from the corner frequency in the
power spectral density of the force measurements.
As the inner radius was increased, the trap stiffness decreased.
The most significant contribution to the decrease in trap stiffness
was the decrease in beam power as more light was removed from the centre
of the beam.
However, this alone doesn't account for the observed behaviour.
For the motile \ecoli{} used in the experiment, the change in
equilibrium position due to the swimming force must also be considered.
When the beam power changes, the ratio of optical force to swimming
force changes, causing the equilibrium position to change.
The optical force curves are highly non-linear, as shown in
figure~\ref{fig:stiffness}~(b); and small changes to the equilibrium position
can have huge changes on the measured trap stiffness.
Using simulations, we were able to explore the effect of different
ratios of swimming force to optical force, producing the shaded region
shown in figure~\ref{fig:stiffness}~(a).

We also used simulations to explore the effect of trap separation
on optical force and particle orientation.
Figure~\ref{fig:stiffness}~(b--c) show the optical force and orientations
for three different beam configurations: two Gaussian-like beams and an
annular beam.
The simulations show that the choice of trap separation as a ratio of
particle length has a significantly greater impact on trap properties
than the choice between Gaussian or annular
beams for low index rod-shaped particles in water.
A mismatch between the particle length and trap separation leads to a lower
stiffness between the force maxima, as previously observed
by \citet{Catala2017Feb} for cylinders in multiple traps
or \citet{Meissner2018Jun} for elongated particles in a single trap.
The observed trap stiffness for a particle in such a trap will strongly
depend on the particle's motility.
If the particle is non-motile, the equilibrium position will be around
the centre of the force maximum and the stiffness will be approximatly 0.
However, for motile particles, the equilibrium position will be closer to
one of the maxima (depending on the swimming direction) and the stiffness
will be significantly larger.
For this reason, care must be taken when using trap stiffness
measurements to quantify the optical trap quality of motile particles.
Further, if the spacing between the two traps was too small, particles
may not sit horizontally, as shown in figure~\ref{fig:stiffness}~(c).
For systems without axial force detection, accurate motility force
(and stiffness) measurements require precise orientation or
accurate estimates for the misalignment.

\subsection{Achieving Fast Orientation using Discrete Potentials}

Using the SLM we were able to gradually change the position of
the two annular beams.
We would like to determine how fast we can change between patterns
and how many intermediate patterns we need in order to change the
orientation of the particle.
To determine this, we simulated a motile \ecoli{} shaped particle.
We assumed the particle swims in one direction with a constant
swimming force.
To quantify the stability of a particular transition, we calculated
the average trap stiffness over the time $\Delta t$ it took the particle to make
the transition in the direction of the desired equilibrium.
This average is given by
\begin{equation}
    \langle k \rangle = \frac{\int_0^{\Delta t} \vec{k} \cdot \vec{x} \text{d}t}{\Delta t}
\end{equation}
where $\vec{k}$ is the trap stiffness
\begin{equation}
    \vec{k} = \frac{\text{d}\vec{F}}{\text{d}\vec{x}},
\end{equation}
and $\vec{x}$ is a unit vector in the direction of the target equilibrium.
This approach assumes that the particles trajectory doesn't vary significantly
from the equilibrium trajectory when Brownian motion is included.
If the particle remains trapped and only rotates, i.e., the position remains
relatively unchanged throughout the trajectory, we can approximate the
above integral by
\begin{equation}
    \langle k \rangle \approx \frac{\int_0^{\Delta t} k_\theta \text{d}t}{\Delta t}
    \label{eq:simple-average-stiffness}
\end{equation}
where $k_\theta$ is the angular trap stiffness
\begin{equation}
    k_\theta = \frac{\text{d}\tau_\theta}{\text{d}\theta}
\end{equation}
for the torque $\tau_\theta$ about the rotation axis $\theta$.

There are multiple sensible values for $\Delta t$ including:
the time required
for the particle to reach an equilibrium orientation, a step size related
to the update rate of the device displaying the pattern, or an arbitrary
duration.
If an arbitrarily long duration is chosen, this will bias $\langle k \rangle$
towards transitions with high trap stiffness at the equilibrium position.
This could be useful for designing trajectories with very stable
intermediate positions.
If instead the rate is equal to the time taken for the
particle to come to equilibrium, $\langle k \rangle$ will be less biased
towards the stiffness at the equilibrium.
In this case, subsequent transitions should be made as soon as the
equilibrium is reached since information about stability at the
equilibrium is not included in the estimate.
Alternatively, if the particle re-orientates much faster than $\Delta t$,
it should be possible to reformulate the above expressions in terms of the
position traversed by the particle before it reaches equilibrium.
The reformulation in terms of position only works for slowly varying fields.
For rapidly varying fields, such as scanned beam optical traps \citep{Carmon2011Jan},
it is necessary to consider the different times scales of the particle
motion, trap frequency and damping.
Here we use the time formulation.

\begin{figure}
    \centering
    \includegraphics[width=\textwidth]{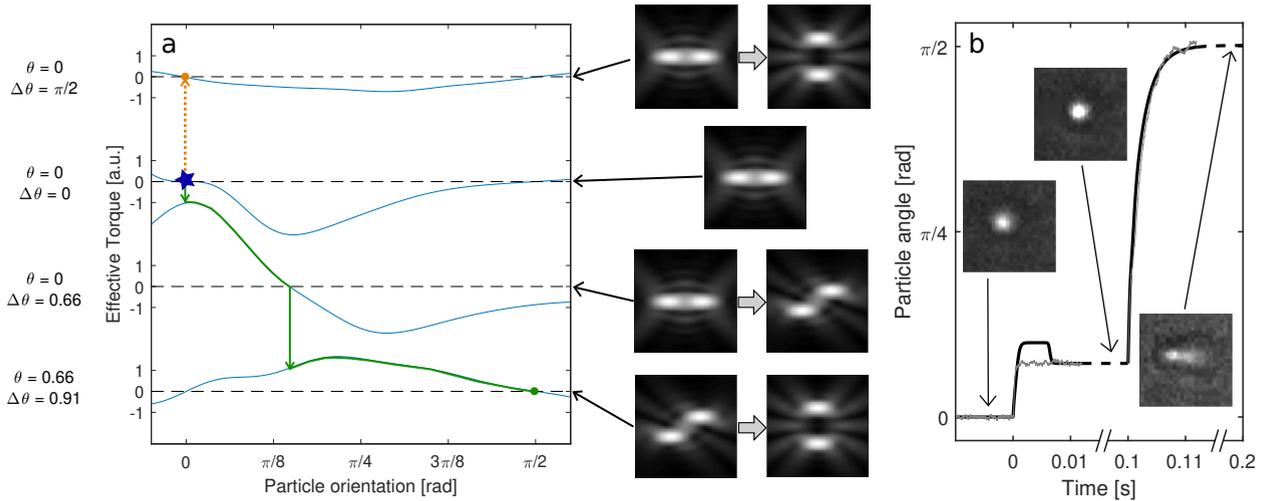}
    \caption{(a) two different beam
    sequences intended to orientate the particle horizontally.
    Starting at the (blue) star with the particle aligned along the
    beam axis, switching directly to two radially displaced annular
    beams (orange dotted line), leads to the particle either escaping
    or aligning vertically in one of the beams.  By adding a single
    intermediate potential the particle can be orientated
    horizontally (solid green line).
    (b) Simulated particle trajectory with one intermediate
    pattern between vertical and horizontal with a delay of 0.1s.
    Solid line shows simulation without Brownian motion,
    jagged gray line shows simulation with Brownian motion.
    Insets show digitally enhanced
    experimental images of an \ecoli{} a short period after each transition.}
    \label{fig:results}
\end{figure}

We choose a moderately long value for $\Delta t$ so that the particle
could come to equilibrium and be stably trapped after each transition.
For a device, such as a liquid crystal based SLM, with a relatively
slow finite update rate (compared to the particle rotation rate),
it is important to have a stable trap position between each orientation
step.
To understand how different step sizes affect the particle orientation,
we simulated the particle swimming in a low Reynolds number environment
in the absence of Brownian motion \citep{Volpe2013Feb}.
By calculating the trap stiffness towards the horizontal equilibrium, we
were able to estimate the strength of the optical trap for different step
sizes, given a particular starting angle.
In our simulations, we observed that the ability to rotate the particle is
affected most by the choice of the first few steps.
As figure~\ref{fig:results}~(a) illustrates, it is very difficult to rotate the
particle from vertical to horizontal with no intermediate steps,
i.e., the particle either falls out of the trap
or gets trapped aligned with the beam axis in one of the annular beams.
With just one intermediate step, it is possible to bypass the
initial low stiffness region and jump to a region where a subsequent
step can correctly orientate the particle.
To understand this behaviour, we plotted the average stiffness towards
the equilibrium (Eq. \ref{eq:simple-average-stiffness}) for a range of
different starting positions and step sizes for our specific annular beam,
see Appendix \ref{sec:appendix}.
For large steps at small initial angles, the average trap stiffness
is small or positive and the resulting angle after the step
doesn't match the trajectory with small steps.
We found the size of the positive stiffness region depends strongly on
the beam properties and the particle shape.
However, a consistent choice for rod shaped particles in Gaussian and annular
beams seemed to be a relatively large initial step ($\sim{}0.15$--$0.7$ radians)
followed by one or more steps to bring the particle to the horizontal equilibrium.

Our simulations suggested it should be possible to orientate the
particle from vertical to horizontal with one intermediate step.
A simulated trajectory (both with and without Brownian motion)
is shown in figure~\ref{fig:results}~b.
When the trap positions are changed, there is a short period of time
before the particle position stabilises.
During this time, the particle also moves vertically/horizontally
in the trap and in some cases this will cause the angle to initially
jump up (as is the case around Time$ = 0$s).
When Brownian motion is included, the size of these
jumps is noticeably reduced in the particle orientation
plot, however the jumps are still observed in the particle position.
To verify our results, we were able to demonstrate orientation experimentally,
the insets in figure~\ref{fig:results}~b show images from the experiment.
The final inset shows the particle aligned approximatly horizontal,
fluctuations in the particle angle are likely caused by Brownian motion.
Our simulations showed that the time for the particle to reach equilibrium
after each step was relatively short.
Movie~2 shows an \ecoli{} being orientated between horizontal and vertical
with two different switching rates.

The method we described here is not specific to the use of annular
beams to orientate particles.
In this paper we choose to focus our attention on annular beams
to create a more consistent narrative; however, we found that a similar
result can be found using more Gaussian-like beams for particle orientation.
The method for quantifying translation and rotation rates
could be extended to other types of transitions; however, the method
rapidly becomes more complex.
For instance, reducing the step size below the time required for the
particle to reach equilibrium makes each trajectory unique.
In this case, the visualisation in figure~\ref{fig:results}~(b) becomes
multi-dimensional and much harder to interpret.
Adding  rotational or translational degrees of
freedom has a similar effect on increasing the dimensionality of the
problem.
The same orientation can be achieved by taking different trajectories
through the translation/orientation hyperspace, each trajectory adding
additional complexity to the problem.
Instead of visualising the effect of each transition,
$\langle k \rangle$ could be used to create an objective function
representing the quality of each trajectory.
This would be useful for planning trajectories in complex light fields,
or for classifying trajectories in fields inducing
rotations or translations, such as for beams with orbital angular momentum.
In particular, this approach could yield interesting results when
using complex light fields to probe the change in a cell's visco-elastic
properties in response to stimuli.


\section{Summary}

Optical tweezers are a useful tool for studying many different
particle shapes including both spherical and non-spherical,
as well as particles which are motile and non-motile.
In order to study, for instance, how a bacterium swims at different
heights above a surface, it is important to not just position
these particles, but also orientate them.
Non-motile particles can be orientated by dynamically varying
the trap power; however, motile particles may escape if the
power is lowered and the trap strength becomes too weak.
By dynamically varying the potential, such as by using a spatial
light modulator to change between discrete beam shapes,
we can change the particle orientation.
For \ecoli{}, we demonstrated this technique by gradually
rotating two annular traps, initially aligned along the beam axis,
until finally aligned transverse to the beam axis.
By simulating the particle and calculating the trap stiffness
along the particle's trajectory, we determined that it should
be feasible, and subsequently experimentally verified,
that the particle orientation can be changed with a single
intermediate step.

In our experiment we used annular beams.
We were motivated to use annular beams as a possible method to
improve angular trap strength for elongated particles, in a
similar way as how annular beams improve axial trap strength
for spherical particles.
Using simulations, we explored how the inner radius of the
annular beam affects the angular trap depth and trap stiffness.
For high refractive rod-shaped particles, we noticed
a significant improvement to the range of particles which
could be trapped when using annular beams.
However, for low refractive index particles, there was very
little improvement.
This suggested that for trapping of \ecoli{} in water, annular
beams wouldn't significantly improve trapping.
However, for trapping higher contrast particles,
such as elongated bacteria or viruses in air or vacuum,
using an annular beam may be beneficial.
We also explored adding spherical aberration to the beam.
In some cases, the shape of the near-field intensity was
less distorted with annular beams,
leading to greater angular trap depth.

\section*{Funding}
This research was funded by the Australian Government through the Australian Research
Council's Discovery Projects funding scheme (project DP180101002).
Isaac Lenton acknowledges support from the Australian Government RTP Scholarship.

\section*{Acknowledgements}
We would like to thank Kate Peters, Steven Hancock, Minh-Duy Phan
and Alvin Lo from the School of Chemistry
and Molecular Biosciences at The University of Queensland
for preparing the \ecoli{} used in these experiments.
We would also like to thank Shu Zhang for assisting with early
experiments using scanned beams; and Carter Fairhall for his work
testing different holographic optical traps.

\appendix
\section{Trajectory average stiffness plots}
\label{sec:appendix}

\begin{figure}
    \centering
    \includegraphics{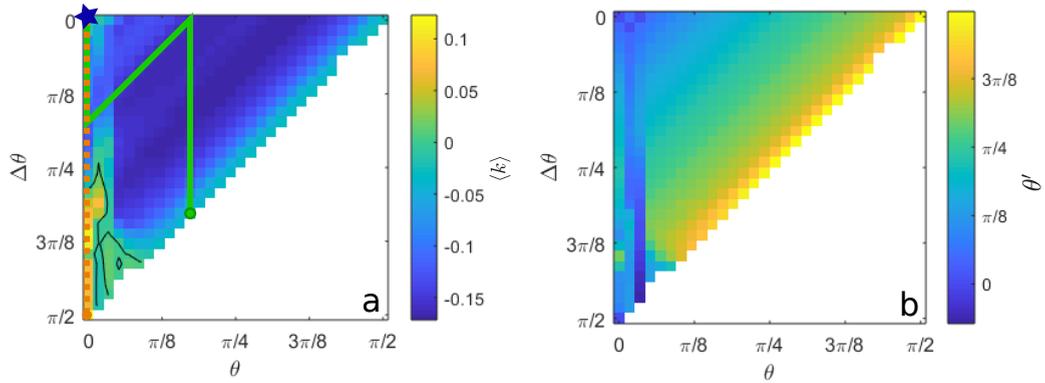}
    \caption{Rotational average trap stiffness and angle calculations for
    rod-shaped particle in an annular beam.
    (a) shows a graphical representation of the average trap
    stiffness towards the equilibrium $\langle k \rangle$ (Eq. \ref{eq:simple-average-stiffness})
    for different step sizes ($\Delta\theta$)
    from a given starting location ($\theta$).
    The orange and green lines show the trajectory depicted
    in figure~\ref{fig:results}~(a).
    (b) shows the resulting particle angle after the step.}
    \label{fig:sup-results}
\end{figure}

Figure~\ref{fig:sup-results} was used to choose the trajectories
shown in figure~\ref{fig:results}.
Figure~\ref{fig:sup-results}~(a) shows the stability of different
steps with respect to the desired final orientation.
For this case, the desired final orientation was
$\pi/2$ radians with respect to the beam axis.
The figure shows each possible beam step size $\Delta \theta$
for the initial orientations $\theta$ between angles
0 and $\pi/2$ radians.
Figure~\ref{fig:sup-results}~(b) shows the corresponding angle
after each step.
These figures suggest that once the particle has rotated past
$\approx \pi/8$ radians, any further rotation of the beams
will lead to the particle moving towards the desired orientation.

In order to rotate the particle past the $\pi/8$ angle, an initial
step or steps need to be chosen.
Figure~\ref{fig:sup-results}~(a) shows that initial steps
larger than approximatly $\pi/4$ radians result in an average
tarp stiffness with the opposite sign.
That is, the particle will not be attracted to the desired
equilibrium.
If we take a small step, the particle will orientate towards
the desired direction, however the range of good choices for
subsequent steps reduces (i.e., the range of positive
$\langle k \rangle$ values becomes larger).
By choosing a step larger than $\pi/8$ radians but less than
the unstable region, we can ensure that the next step is able
to orientate the particle in the desired direction.

This particular configuration resulted in the particle angle
after a stable step being equal to the particle angle with
many infinitesimally small steps ($\Delta\theta = 0$ line in
figure~\ref{fig:sup-results}~(b)).
It is conceivable that there may be some situations where the
particle angle after a transition doesn't match the infinitesimally
small step case.
In these cases, it might be more helpful to plot the difference
between the resulting particle angle and the target
(infintesimally small step) angle instead of
figure~\ref{fig:sup-results}~(b).
It may be necessary to generate multiple versions
of these plots for each case where the angles don't match,
or to generate a 3-dimensional
plot for $\langle k \rangle$ and $\theta'$ with all combinations
of $\theta$, $\Delta\theta$ and initial particle orientation.
Alternatively, it may be easier to simply avoid such cases
and only choose transitions which lead to the particle having
the same orientation as the infinitesimally small step case.

\printcredits

\bibliographystyle{cas-model2-names}

\bibliography{paper}

\clearpage


\bio{isaac}
Isaac Lenton is a PhD student under the supervision of
Timo Nieminen, Alex Stilgoe and Halina Rubinsztein-Dunlop.
His PhD topic is on theory, modelling and control of optical
tweezers, with a focus on studying motile organisms using
optical tweezers.
However, he is also working on a range of other projects
tangentially related to his PhD topic including light shaping,
imaging using spatial light modulators,
machine learning, and fabrication of micro-fluidic devices.
\endbio
\vskip20pt

\bio{declan}
Declan Armstrong is a PhD student working under the supervision of Halina Rubinsztein-Dunlop. His PhD topic is focused on modelling and testing theories of biohydrodynamics,
with particular focus on the study of cells under propulsion. Further areas of interest expand to studying the behaviour and control of optically trapped biological cells, and particles undergoing stochastic motion.
\endbio
\vskip20pt

\bio{alex}
Dr Alexander Stilgoe received his PhD from The University of
Queensland in 2012.
His recent work has been focused on understanding light scattering
for applications to imaging and force measurement and the dynamics and interactions of particles and molecules on the microscopic and nanoscopic scale.
He has performed a large amount of computational work and some experimental work on the interaction of complex electromagnetic fields with microscopic particles, e.g. multiple beams and structured light under different conditions.
\endbio
\vskip10pt

\bio{timo}
Timo Nieminen obtained his PhD at The University of Queensland in 1996.
His main research interests can be broadly classified as electromagnetic
theory and computational electromagnetics. The main focus of this work
is the theory and modelling of optical trapping, and the scattering of
electromagnetic waves by particles. This work includes the development
of the Optical Tweezers Toolbox, a freely available Matlab toolbox for
the computational modelling of optical tweezers, which is available at
\url{https://github.com/ilent2/ott}.
Other interests include physics education, the history of physics and
the physics of history, photonics, biological applications of optics,
astrophysics, and the cross-disciplinary application of research
methodology and tools.
\endbio
\vskip-10pt

\bio{halina}
Professor Halina Rubinsztein-Dunlop's research interests are
in the fields of atom optics, laser micromanipulation, nano
optics, quantum computing and biophotonics.
She has long standing experience with lasers, linear and nonlinear
high-resolution spectroscopy, laser micromanipulation, and atom
cooling and trapping. She was one of the originators of the
widely used laser enhanced ionisation spectroscopy technique
and is well known for her recent work in laser micromanipulation.
She has been also working (Nanotechnology Laboratory, G\"oteborg,
Sweden) in the field of nano- and microfabrication in order
to produce the microstructures needed for optically driven
micromachines and tips for the scanning force microscopy
with an optically trapped stylus.
\endbio

\end{document}